\documentclass[letterpaper, aps, tightenlines, nofootinbib, 11pt]{article}
\pdfoutput = 1

\usepackage{setspace}
\usepackage[bookmarks = false, colorlinks = true, linkcolor = blue, citecolor = purple]{hyperref}
\usepackage{shortcuts}
\usepackage{titlesec}
\usepackage{cite}
\usepackage{tocloft}
\usepackage[vcentermath]{youngtab}
\usepackage{tensor}

\usepackage{tikz}
\usetikzlibrary{arrows,decorations.pathmorphing,backgrounds,positioning,fit,petri}

\usepackage[margin = 2.5cm]{geometry}
\begin{document}

\begin{center}

\thispagestyle{empty}

\vspace*{5em}

{\LARGE \bf Quantum Mechanics in the Infrared}

\vspace{1cm}

{\large \DJ or\dj e Radi\v cevi\'c}
\vspace{1em}

{\it Stanford Institute for Theoretical Physics and Department of Physics\\ Stanford University \\
Stanford, CA 94305-4060, USA}\\
\vspace{1em}
\texttt{djordje@stanford.edu}\\

\vspace{0.08\textheight}
\begin{abstract}
This paper presents an algebraic formulation of the renormalization group flow in quantum mechanics on flat target spaces. For any interacting quantum mechanical theory, the fixed point of this flow is a theory of classical probability, not a different effective quantum mechanics. Each energy eigenstate of the UV Hamiltonian flows to a probability distribution whose entropy is a natural diagnostic of quantum ergodicity of the original state. These conclusions are supported by various examples worked out in some detail.
\end{abstract}
\end{center}

\newpage


\section{Introduction}

Quantum chaos is an old topic motivated by a simple question: what is the analogue of classical chaos in quantum systems? Equivalently, how are apparent randomness of motion and sensitivity to initial conditions encoded in quantum dynamics at long times? This question has appeared in many guises, e.g.~in the study of the spectrum of highly excited states of quantum systems \cite{Dyson:1970tza, Gutzwiller:1990, Wigner:1951, Berry:1989}, in connection to the localization/thermalization transition in many-body systems (see e.g.~\cite{Nandkishore:2014kca}), in attempts to understand the black hole information paradox \cite{Maldacena:2015waa, Shenker:2013pqa}, and even in relation to reformulating and proving the Riemann hypothesis \cite{Berry:1986}.

Ultimately, the study of quantum chaos reduces to understanding the long-time (but not necessarily long-distance) behavior of a quantum system. For instance, spectral statistics of a Hamiltonian can be probed using correlation functions at long times, or, equivalently, by calculating a certain path integral over long classical trajectories that fill up the available phase space \cite{Gutzwiller:1971, Berry:1977, Bogomolny:1988, Muller:2005, Sieber:2001}. By now there exists a wealth of information about the long-time regime of many quantum systems, and many universal properties that diagnose quantum chaos (or lack thereof) are known. In particular, correlators at the longest time scales available in the system govern spacings between nearby eigenvalues of the Hamiltonian, and these spacings are universally captured by simple random matrix theory in all chaotic systems \cite{Berry:1989, Bohigas:1983er, Muller:2009}.

Given the overwhelming evidence for long-time universality, and more generally given the importance of studying quantum chaos in connection to questions in both the condensed matter and the high energy communities, developing a systematic renormalization group (RG) procedure that takes us to long times would be of great interest. This cannot be the usual RG in the sense of Wilson and Kadanoff \cite{Kadanoff:1966wm, Wilson:1971bg}. For instance, the desired procedure should work in quantum mechanics, i.e.~in $0+1$ dimensions, where there is no analogue of ``spin-blocking'' available. Another reason is that  this procedure should also work for conformal field theories, which are fixed points under Wilsonian RG.

In this paper we will formulate long-time RG for quantum mechanics (QM). There are no spatial dimensions here, so it is possible to think of QM as a QFT in $0 + 1$ dimensions and apply the usual Wilsonian ideas to the path integral in order to orient ourselves; we will later comment on why they are not enough. Let us consider a QM theory given by
\bel{
  \int [\d q] \exp\left\{i \int \d t \left[ \left(\pder q t\right)^2 - V(q) \right]   \right\}.
}
The field $q$ has mass dimension $-1/2$ and it takes values on a compact target space.\footnote{In order to meaningfully talk about chaos, the phase space has to have finite volume. This is why we only work with compact target spaces in this paper.} The linear size $\sigma$ of the target is a dimensionful coupling in the theory. Under Wilsonian RG, $\sigma$ will flow towards zero, and the theory will become strongly coupled. More precisely, the dimensionless parameter that sets the strength of interactions is $\sigma^2/\tau$, where $\tau$ is the time scale of interest. For fixed $\sigma$, short times $\tau$ are described by coherent, classical dynamics, while long times represent the limit in which quantum fluctuations are important.\footnote{Note that we set $\hbar = 1$ as in any other QFT. Literature on quantum chaos often studies the semiclassical limit $\hbar \rar 0$ while keeping the target space fixed, but we follow the field-theoretic practice of keeping $\hbar = 1$ and using other parameters --- in this case $\sigma^2/\tau$ --- as dimensionless couplings that govern quantum fluctuations.
\\
\indent It is also important to contrast the two kinds of classical regimes one can talk about. At $\sigma^2/\tau \rar \infty$, the path integral is dominated by trajectories around a unique classical saddle point; the kinetic term controls most of the dynamics and there is typically ballistic transport with very little diffusion. At $\sigma^2/\tau \rar 0$, the path integral receives contributions from all possible long classical trajectories/saddle points. The conventional lore is that this regime reflects the statistics of energy eigenstates, as these are the only states with trivial time evolution --- it is this property that allows them to probe long times without getting averaged to death. Looking at the Hamiltonian $-\frac{\hbar^2}{2\mu} \pdder{}q + V(q)$, one sees that studying the spectrum at $\hbar \rar 0$ makes states with high momenta $(\sim 1/\hbar)$ have finite energy, thereby allowing us to study them in a controlled way. This way the classical limit gives us access to the long-time dynamics.}

Wilsonian RG has very limited use here. QM dynamics is quasiperiodic whenever there are at least two incommensurate frequencies in the problem, so unless the theory is free or otherwise fine-tuned, the usual RG equations will be oscillatory and will not reduce long-time dynamics to some universal fixed point \cite{Polonyi:1994pn}. Another way to see this problem is to notice that the IR is not easily expressed as an effective field theory because all powers of $q$ are, na\"ively, relevant operators; an explicit regularization is needed to make sense of the IR. This issue is related to the fact that approximate conformal symmetry in QM only arises in conjunction with singular potentials and other pathologies \cite{deAlfaro:1976vlx, Chamon:2011xk} (see also \cite{Maldacena:2016hyu, Polchinski:2016xgd} for new results on the dual nearly-AdS$_2$ space).

In this paper we avoid path integral methods and present a very natural way to describe a monotonic RG flow in the Hamiltonian formalism. The idea is very much in the spirit of Wilsonian RG: at each step, we reduce the algebra of observables to an appropriately chosen subalgebra. (Usual Kadanoff decimation can be expressed in this language.) As we will argue in detail in Section 2, at short times this algebraic coarse-graining naturally corresponds to changing the cutoff used to discretize time in the definition of the QM path integral. The true benefit of our framework, however, is that it allows us to rather transparently understand what happens in the deep IR, i.e.~when the time step becomes finite, or even when it is of the order of the recurrence time. The upshot is that going to longer time scales (or smaller algebras) does not take us to a different, effective QM; rather, each decimation turns some degrees of freedom into classical random variables, and flowing to the deep IR turns a QM theory into a theory of classical probability whose event space corresponds to the remaining Abelian algebra of observables. This is decoherence.

This conclusion has some far-reaching implications. In particular, if it is assumed that all well-defined RG procedures can be defined algebraically, then the work presented here implies that there is essentially \emph{no} effective, UV-independent, quantum description of interacting IR dynamics in $0 + 1$ dimensions. The only exception is a trivial one: time-independent states in a given theory will have well-defined IR behavior. This means that there is no universal quantum-mechanical fixed point in \emph{any} model, and the only sensible fixed points in QM are time-independent classical probability distributions that come from decimating energy eigenstates in the UV. For models whose target spaces are Abelian groups, we will put these claims on solid footing by the end of Section 2.

In Sections 3 and 4 we study algebraic coarse-graining of energy eigenstates and its connection to diagnosing chaos. We restrict ourselves to flat target spaces with various potentials. In each case we decimate the algebra of observables, and we follow how eigenstates of the Hamiltonian decohere and become mixed following a very specific pattern governed by the center of the subalgebra. Along the way we note that the von Neumann entropy of the resulting mixed states acts both as a Zamolodchikov $c$-function and as a measure of quantum ergodicity or ``eigenstate thermalization,'' i.e.~as a measure of how randomly distributed the eigenstates are compared to those of a free Hamiltonian that is naturally defined on the target group.

In the Conclusion, we emphasize the lessons learned and offer more comments on related topics. In particular, we point out other systems of potential interest that this procedure extends to. As a bonus, we also clarify the nature of  the Hilbert space of a single Majorana fermion.

\section{Coarse-graining algebras}

\subsection{Setup}

We will exclusively work with QM theories whose target $G$ is a $d$-torus or, when regulated, a discrete group of the form $(\Z_N)^d$. The generalization of our results to other (nonabelian) compact groups is straightforward but often tedious, and we will briefly review how it is done at the end of this Section.

The operator algebra inherits the group structure of the target space in the following way. Let $g_i$ be the $d$ generators of $G$. There are then $2d$ generators of the algebra. These are the $d$ position operators, $U_i$, and the $d$ shift or momentum operators, $L_i$. The Hilbert space appropriate to the given target group $G$ is spanned by a set of vectors $\{\qvec{g_1^{n_1}\cdots g_d^{n_d}}\}$ labeled by elements $g = \prod_i g_i^{n_i} \in G$ with $n_i = 0, \ldots, N - 1$. (For continuous $G$, we will use $e^{i \theta_i \b e^i}$ with $\theta_i \in [0,\,2\pi)$ instead.)  The position operators are chosen such that each of these basis states is an eigenstate of all the $U_i$'s; the spectrum of $U_i$ is then given by a unitary representation of $U(1)$ or $\Z_N$. The shift operators in the discrete case are chosen so that $L_i \qvec g = \qvec{g_i g}$; in the continuous case, $L_i = \exp\left\{i\, \d \theta^i \pder{}{\theta_i}\right\}$. Since $G$ is Abelian, all shift generators commute with each other, and the only nontrivial commutation relation is
\bel{\label{eq comm}
  U_i L_i = g_i L_i U_i.
}

For simplicity, we now specialize to $d = 1$. Consider first the continuous case: a particle on a circle. Let us focus on propagation over a very short time interval $\d t$ (relative to any other time scale in the problem). The length $\d t$ can be thought of as a regulator in the path integral, which is formally defined as a product of $\sim\frac{1}{\d t}$ ordinary integrals. If the kinetic term has two derivatives, any two trajectories that differ by less than $\sqrt{\d t}$ at all times will have the same action and will therefore constructively interfere in the path integral. Higher derivative terms in the action will merely change the power of $\d t$ without changing the qualitative picture.

We can thus conclude that a temporal cutoff induces a target space cutoff, in the sense that a QM path integral whose time is measured in steps of $\d t \rar 0$ is well-approximated by $\frac 1{\d t}$ sums over the target space with step $\sqrt{\d t}$. Conversely, a target space cutoff $\d x$ induces a natural time step $\d x^2$, the time it takes to diffuse across length $\d x$; any trajectories that started within length $\d x$ of each other will be mixed together after time $\d x^2$. However, if $\d x \sim 1$ as in, say, $G = \Z_2$, then this heuristic makes no sense, as $\d t$ is no longer infinitesimal.

Now we see in more detail how Wilsonian RG leads to trouble. As we begin the flow in a theory with target $U(1) = \lim_{N \rar \infty} \Z_N$, the natural time step is $\d t \sim 1/N^2 \rar 0$. As $\d t$ is increased, $N$ will decrease, and at some point it will stop being the largest quantity in the problem. In fact, after sufficiently many na\"ive Wilsonian decimations, $N$ will become finite, and at that point the natural time step is finite and we have no right to view the effective theory as being given by a path integral.

This means that we need a better framework for understanding what happens after many decimations. The connection between $\d t$ and $\d x$ (or $N$) provides a natural way forward in a Hamiltonian framework: one flows to the IR by coarse-graining the target space without any explicit reference to time. We will henceforth refer to the starting target space as the UV, and to the final target space as the IR.  In order to flow from the UV while maintaining all the IR correlations, we follow the same strategy that proved useful in understanding entanglement in general QFTs \cite{Radicevic:2014kqa, Casini:2013rba, Ghosh:2015iwa}: we construct a sequence of operator algebras, each contained in the previous one, and we choose that the position generator in each new algebra can distinguish fewer different positions. For instance, if $U$ was a position generator in the UV $\Z_N$ theory, $U^2$ will be a position generator appropriate for a $\Z_{N/2}$ theory, $U^4$ will be appropriate for a $\Z_{N/4}$ theory, and so on. States --- represented by density matrices that belong to algebras --- are then naturally projected along the flow in such a way as to ensure that expectations of all IR operators are preserved.

In the next subsection we will present this construction in detail, but here we highlight how it differs from Wilsonian RG as we typically understand it:

\begin{enumerate}
  \item We flow to longer time scales, but not necessarily to smaller energies. However, the long-time dynamics encodes information about gaps between nearby eigenvalues in the Hamiltonian, while dynamics at shorter times encodes coarser properties of the spectrum. In this (admittedly vague) sense we are flowing towards smaller energy scales.
  \item States in the IR will typically be mixed, even if we started from a pure state in the UV. In QFT we work with the pure state sector in the IR, but here we will not be able to do this because the UV and IR sectors are strongly coupled. The coupling between the UV and IR in a QFT is governed by the small width of the momentum shell being integrated out; here we have no such parameter. Thus, evolution of a pure IR state will almost always make it mixed.
  \item In a similar vein, time evolution in the IR is not governed by an effective Hamiltonian built out of IR operators. The mixing of states is described by a classical probability distribution whose evolution is set by the time-dependent expectation values of the UV operators in the Hamiltonian.
  \item The Kadanoff decimation in QFT can also be represented as a reduction in algebras, as we are there removing position and shift operators associated to modes with high spatial momentum. In QM we have no sense of spatial momentum, and instead we are removing position operators while leaving the shift operators intact. Unlike in QFT, our IR algebras will always have a nontrivial center. The superselection sectors labeled by the center eigenvalues are the sectors that are being mixed; center operators are the classical variables. After many decimations, we will be left only with the center. This is now a wholly classical regime, with each density matrix diagonal in the center eigenbasis.
  \item There exists scheme dependence. For instance, coarse-graining shift operators instead of the position operators leads to a different IR with different center operators. For theories with a quadratic kinetic term and a position-dependent potential, the scheme in this paper is natural.
  \item We will be particularly interested in the IR fixed point in which we have no more position operators to remove (the ``deep IR''). In principle, we can then continue coarse-graining the shift operator algebra. This leads to a trivial algebra that contains just the identity, which is the ultimate fixed point of our RG. However, in this paper we focus on the fixed point of the scheme in which just the position generators are coarse-grained.
  \item The dimension of the Hilbert space never changes in our setup, but all density matrices become block-diagonal in the center eigenbasis. The labels of these sectors are classical variables, and so one can say that the effective amount of quantum variables in the theory decreases after each coarse-graining.
\end{enumerate}

\subsection{The decimation procedure}

Let us now describe in some detail the proposed target space coarse-graining in a $\Z_N$ theory. For simplicity, we take $N$ to be a power of two. The starting algebra of observables $\A_0$ consists of all possible products of position and shift generators $U$ and $L$, with $U^N = L^N = \1$. In position basis, $U = \trm{diag}(g^n)$ where $g = e^{2\pi i/N}$ and $n = 0, \ldots, N - 1$. We define a decimation to be the map $\A_r \mapsto \A_{r + 1}$, where $\A_{r + 1}$ is generated by the same shift operator as $\A_r$ and by the square of the position generator of $\A_r$. Since we are starting from $\A_0$, this means that $\A_r$ is generated by $U^{2^r}$ and $L$. The center of $\A_r$ is the Abelian group generated by $L_r \equiv L^{N/2^r}$, as can be easily checked by using the commutation relation \eqref{eq comm}.

Whenever we represent elements of $\A_r$ as matrices on the $N$-dimensional Hilbert space (which is the same for every $r$), we will always work in a basis that diagonalizes $L_r$. All elements of $\A_r$ will be block-diagonal in this basis. We will refer to these blocks as superselection sectors, and we will index them with $k = 0, \ldots, 2^r - 1$ such that the eigenvalue of $L_r$ in sector $k$ is $g_r^k \equiv g^{N k / 2^r}$.

Any density matrix $\rho$ is an element of the algebra of observables. It can be represented as
\bel{
  \rho = \sum_{\O \in \A} \rho_\O \O.
}
In our setup, it can be shown \cite{Radicevic:2016tlt} that
\bel{
  \rho_\O = \frac1N \avg{\O^{-1}}.
}
Decimation also maps $\rho_r \mapsto \rho_{r + 1}$ such that $\Tr(\rho_r \O) = \Tr(\rho_{r + 1} \O)$ for all $\O \in \A_{r + 1} \subset \A_r$. This decimation is simple in the eigenbasis of $L$: one just restricts $\rho_r$ to the block-diagonal form with $2^{r + 1}$ blocks, and sets all the other entries to zero. Only eigenstates of $L_{r + 1}$ remain pure after this decimation.

It is instructive to see how certain states map. Let us take as an example the $\Z_4$ theory. After one decimation, the position eigenstates $\qvec{\b 1} = \qvec{g^0}$ and $\qvec{g^2}$ both become uniform mixes of $\frac1{\sqrt 2}\left(\qvec{\b 1} + \qvec{g^2}\right)$ and $\frac1{\sqrt 2}\left(\qvec{\b 1} - \qvec{g^2}\right)$. On the other hand, the $L_1$ eigenstates $\frac1{\sqrt 2}\left(\qvec{\b 1} \pm \qvec{g^2}\right)$ remain pure. The same thing happens for the pair of states $\qvec g$ and $\qvec{g^3}$. In general, any linear combination of pure nondegenerate eigenstates of $L_r$ will become mixed, i.e.
\bel{
  \rho_0 = \sum_{k,\,l} \alpha_k \alpha_l^* \qvec{\psi_k}\qvecconj{\psi_l} \ \mapsto\ \rho_r = \bigoplus_k |\alpha_k|^2 \qvec{\psi_k}\qvecconj{\psi_k}.
}
for any set $\{\qvec{\psi_k}\}$ of nondegenerate eigenstates of $L_r$.

Going back to the $\Z_4$ example, we may thus view the decimated system as a coupled set of two Hilbert spaces (sectors) of dimension two, or of two $\Z_2$ theories. Each sector is spanned by states with the same $L_1$ eigenvalue; e.g.~one sector is spanned by $\frac1{\sqrt 2}\left(\qvec{\b 1} + \qvec{g^2}\right)$ and $\frac1{\sqrt 2}\left(\qvec{g} + \qvec{g^3}\right)$. Each sector can thus be viewed as a $\Z_2$ theory with a prescribed winding along this decimated group manifold.

As another example, consider the $U(1)$ theory (regulated to look like $\Z_N$ with very large $N$) and perform one decimation. Any state in the original theory will now be a mix of two states living on circles half the size of the original one; one of these two states will have trivial winding around the new circle, while the other state will pick up a factor of $g_1 = g^{N/2} = -1$ when winding. After $r$ decimations, each state in the original theory will become a mix of $2^r$ states, each of which with a definite winding $g_r^k = e^{2\pi i k/2^r}$ around the decimated target space.

The reduced density matrix after $r$ decimations thus takes the form
\bel{
  \rho_r = \bigoplus_{k = 0}^{2^r - 1}\, p_k \,\varrho_k.
}
Here, $\varrho_k$ are the $2^r$ diagonal blocks of the original density matrix in the $L$ eigenbasis, normalized so that each has unit trace in its $\frac N{2^r}$-dimensional subspace. The normalization factors $p_k$ form a probability distribution on the space of sectors. They are given by the expectations of the center of $\A_r$, and it is straightforward to check that
\bel{\label{eq pk vev}
  p_k = \frac1{2^r} \sum_{l = 0}^{2^r - 1} \avg{L_r^l} g_r^{-k l}.
}
Another useful way to express these probabilities is via the eigenstates $\qvec \ell$ of the shift operator $L$. These states satisfy $L\qvec \ell = g^\ell \qvec \ell$, $\ell = 0, \ldots, N - 1$, and for a given state $\qvec \Psi$ the classical probabilities are
\bel{\label{eq pk overlap}
  p_k = \sum_{\substack{\ell = k + 2^r k'\\ k' = 0,\, 1,\,\ldots,\, \frac N{2^r} - 1}} \left|\qprod {\ell} \Psi\right|^2.
}
We already see from this form that in the deep IR, when $2^r = N$, the sector probabilities reduce to the classical probabilities of eigenstates of operators in the remaining Abelian algebra. We also see that $p_k$ is the momentum-space analogue of Wigner, Husimi, and other quasiprobability distributions that are often the main tools in studying eigenstate statistics, as in \cite{Berry:1977, Bogomolny:1988, Nonnenmacher, Chirikov:1988} and references therein. However, we stress that in our case the states $\ell$ are determined by the group structure of the target space.

For any initial pure state, the matrices $\varrho_k$ all describe pure states.\footnote{An educated reader might guess that the $\varrho_k$ should be thermal states, at least for eigenstates of chaotic theories, as per the eigenstate thermalization hypothesis. This is not the case in quantum mechanics.} The only mixing that happens is between different sectors, and the entropy of this mixing is the Shannon entropy
\bel{
  S_r = -\sum_k p_k \log p_k.
}
This is a nondecreasing function of $r$ for any starting state so it can be identified as a $c$-function for our RG, though it is generally time-dependent. As we will discuss next, there is no time dependence for energy eigenstates of the UV Hamiltonian, and we may define the average IR entropy of the Hamiltonian as
\bel{\label{eq S avg}
  \bar S_r \equiv \frac1N \sum_n S_r^{(n)},
}
where $\qvec n$ are the $N$ energy eigenstates and $S_r^{(n)}$ are their entropies after $r$ decimations. The average IR entropy is bounded by $r \log 2$, and after each decimation it cannot rise by more than $\log 2$. This in turn implies that in the deep IR, at $r = \log_2 N$, the IR entropy is bounded by
\bel{
  \bar S\_{IR} \equiv \bar S_{\log_2 N} \leq \log N.
}
In the examples we will find that this entropy is connected to the mixing properties of the underlying classical system --- or, more precisely, that it measures the quantum ergodicity of the theory.

\subsection{Time evolution in the IR}

In the UV, dynamics is given by $\rho^t = e^{iHt}\, \rho\, e^{-iHt}$. In the IR, things are no longer as simple. If the UV Hamiltonian contains no operators that are lost after $r$ decimations, i.e.~if it is block-diagonal in an $L_r$ eigenbasis with $H = \bigoplus_k H_k$, then the appropriate probability distribution $p_k$ of any state will be time-independent. The time evolution of any state will then just be captured by $\varrho_k^t = e^{iH_k t} \varrho_k e^{-iH_k t}$.

However, as long as the Hamiltonian is not the identity matrix, eventually the decimation will make us lose operators that govern time evolution. Once this happens, no operator in the remaining observable algebra will be able to determine how the sector probabilities evolve, unless the UV state was an energy eigenstate to begin with. This follows easily from eq.~\eqref{eq pk vev}, with
\bel{
  \der{p_k}t = - \frac i{2^r} \sum_l \avg{[H, L_r^l]} g_r^{-k l};
}
any operator in $H$ that is also in $\A_r$ will necessarily commute with $L_r^l$, and hence  $\d p_k/\d t$ will be determined by the UV expectation values alone. From the point of view of the IR, the sectors all couple to each other as they evolve in time, and in addition this evolution has external parameters whose time dependence is specified by the UV state and Hamiltonian.

The above conclusions are very general; we could have started from any other algebraic decimation scheme, and the IR time-dependence would have still been encoded in the UV data. Assuming that algebraic decimation is the only controlled way of performing RG, this means that there is \emph{no} effective IR theory that governs the dynamics of a QM system. Said another way, time evolution in QM is always sensitive to UV physics because of the nontrivial center. The only exception to this rule justifies the lore stated in the Introduction: only eigenstates of the Hamiltonian, due to their trivial time-dependence, have a UV-insensitive time evolution at long time scales. Therefore, the only IR physics that one can meaningfully talk about is obtained by decimating energy eigenstates.


\section{Examples}

We now work out some examples in increasing order of complexity. Our goal is to study the average IR entropy. Where available, we perform computations analytically; otherwise, we perform preliminary numerical studies and leave a more detailed analysis for future work. We will always work with the class of Hamiltonians with a quadratic kinetic term,
\bel{\label{eq H prototype}
  H = \alpha(L + L^{-1}) +  \sum_n \beta_n (U^n + U^{-n})
}
for some constants $\alpha$ and $\beta_n$. Note that the continuum limit is reached by letting $N \rar \infty$ while scaling $\alpha \sim 1/\d \theta^2 \sim N^2$ and $\beta_n \sim N^0$.

\subsection{Free particle on a circle}

Let us start with a free particle moving on a $\Z_N$ group in the limit of large $N$.  The Hamiltonian is
\bel{
  H = \alpha(L + L^{-1}) \stackrel{N \rar \infty}{\longrightarrow} - \frac1{2 \mu} \pdder{}{\theta}.
}
(Note that we drop constant terms in $H$ when going to the continuum.) Eigenstates and eigenvalues of the Hamiltonian are
\bel{
  \psi_n(\theta) = \frac1{\sqrt{2\pi}} e^{i n \theta}, \quad E_n = \frac{n^2}{2 \mu}, \quad n \in \Z.
}
Sector probabilities for each state $\psi_n$ are easily calculated from \eqref{eq pk vev} or \eqref{eq pk overlap}, and they are\footnote{Due to the degeneracy of $\psi_n$ and $\psi_{-n}$, we could have chosen different eigenstates, and this would have led to different sector probabilities. This is not generic; typically we will deal with completely nondegenerate systems. Even if we encounter small degeneracies, they will only affect the IR entropies by finite amounts, and this will not change any of our conclusions.}
\bel{
  p_k = \delta(k + n \equiv 0\, \trm{mod}\,2^r).
}
Being eigenstates of $L_r$, the states $\psi_n(\theta)$ retain their purity during decimation.  For each $n$, there is a unique $k$ such that $n + k \in 2^r \Z$, and this is the sector the state is in. Note that this means that the classical probability is a $\delta$-function for each eigenstate, independent of the number of decimations. The average IR entropy is thus
\bel{
  \bar S\_{IR} = 0.
}
The fact that the average entropy is independent of $N$ is \emph{one} possible signature of quantum ergodicity, as we will discuss later.

The Hamiltonian commutes with all of the decimated operators in this theory, and therefore all the $p_k$'s for all the states will stay constant. In fact, they can be found explicitly for any state $\Psi(\theta) = \frac1{\sqrt{2\pi}}\sum_n \Psi_n e^{in\theta}$. 
Sector probabilities are given by \eqref{eq pk overlap},
\algnl{
  p_k
  &= \sum_{n + k \equiv 0\,\trm{mod}\,2^r} |\Psi_n^2|.
}
Note that once all of the $U$'s are decimated, we get the famous result $p_k = |\Psi_k^2|$. This is the aforementioned decoherence in the IR. Further decimation now coarse-grains the Abelian algebra of shift operators. This can be interpreted as going to ever smaller $\sigma$-algebras of events in classical probability. It is only because of this decimation in the deep IR that the free theory is not, strictly speaking, a fixed point of our procedure.


\subsection{SHO on a circle}

Now we add the simplest possible potential to the free particle:
\bel{\label{eq H sho}
  H = \alpha (L + L^{-1}) + \beta (U + U^{-1}) \stackrel{N \rar \infty}{\longrightarrow} -\frac1{2\mu} \pdder{}\theta + \mu\omega^2 \cos\theta.
}
This is the potential of the quantum rotor. The eigenstates are solutions to the Mathieu equation and can be manipulated numerically.


Before proceeding, let us notice that potentials $U^n + U^{-n}$, for any $n \leq N/2$, all lead to the Schr\"odinger equation of a Mathieu type. At $n = N/2$ we obtain an extreme case that can be solved exactly due to the high degree of symmetry ($L^2$ and all its powers commute with this Hamiltonian). Here the theory is
\bel{
  H = \alpha(L + L^{-1}) + \beta U^{N/2}.
}
To control the computation we must keep thinking about the discrete target space, as the potential oscillates with a wavelength equal to the shortest distance in the problem.

This system splits into $N/2$ two-dimensional sectors that are labeled by eigenvalues of $L^2$. Each sector is spanned by two states, $\qvec{\trm{odd}, n} \propto g^n\qvec{g} + g^{3n} \qvec{g^3} + \ldots$ and $\qvec{\trm{even}, n} \propto \qvec{\1} + g^{2n}\qvec{g^2} + \ldots$ with a sector Hamiltonian
\bel{
  H_n = \bmat \beta {\alpha(g^n + g^{-n})} {\alpha(g^n + g^{-n})} {-\beta}.
}
Here $n$ runs from 0 to $N/2 - 1$. The eigenstates of the full Hamiltonian are thus states
\bel{
  \qvec{n_\pm} \propto \alpha (g^n + g^{-n}) \qvec{\trm{odd},n} + \left( \beta \pm \sqrt{\alpha^2 (g^n + g^{-n})^2 + \beta^2 }\right) \qvec{\trm{even}, n}
}
with eigenvalues
\bel{
  E_{n_\pm} = \pm \sqrt{\alpha^2 (g^n + g^{-n})^2 + \beta^2 }.
}
These are eigenstates of $L^2$ and all other powers of $L$. In particular, these states are eigenstates of $L_r$ up until the last decimation, when $r = \log_2 N$. Hence, until the last decimation, eigenstates stay pure. At the last decimation, the entropy of each eigenstate rises by a term bounded by $\log 2$.



Going back to the $U + U^{-1}$ potential \eqref{eq H sho}, we numerically investigate how periodic Mathieu functions (eigenstates in the continuum limit) behave under decimation. These are functions $\psi^S_n(\theta)$ and $\psi^C_n(\theta)$ that, as $\omega \mu \rar 0$, reduce to ordinary sines and cosines. For a fixed $\omega\mu$, the states with $n \ll \omega\mu$ are oscillations localized at $\theta = \pi$, the minimum of the potential; at $n \gg \omega\mu$, the wavefunctions approach ordinary sines and cosines spread out over the entire circle.

We have enough control over these functions to immediately bound their IR entropy $\bar S_r$ at the maximal number of decimations, $r = N \log 2$. To do this, it is enough to use \eqref{eq pk overlap} and calculate the overlaps of $\psi_n^{S/C}(\theta)$ with the plane waves $\frac1{\sqrt{2\pi}} e^{i \ell \theta}$. At $n \gg \omega\mu$, only two sectors ($\ell = \pm n$) have a nontrivial probability, leading to an entropy of $\log 2$ for each high-momentum state. At $n \lesssim \omega\mu$, there are more sectors with nontrivial probabilities, but they are all located at $|\ell| \lesssim \frac{\omega \mu}n$, and the entropy for these states can be very roughly bounded by $\log \frac{\omega\mu}n$. At $\omega \mu \gg 1$, the contribution from all the low-energy states is then approximately bounded by $\sum_{n = 1}^{\omega \mu} \log \frac{\omega\mu}n \sim \omega \mu$, and the average IR entropy is
\bel{
  \bar S\_{IR}  \lesssim \frac{\omega \mu + (N - \omega \mu) \log 2}N \stackrel{N \rar \infty}\longrightarrow \log 2.
}The average IR entropy is finite in the continuum limit, and approximately satisfies the same bound as the one we analytically derived for the $U^{N/2}$ potential. As before, the finiteness of the IR entropy is a signature of the system not being ergodic.

It is worthwhile pointing out that the IR entropy increases with $\omega\mu$. This increase in entropy comes from the low-energy states that are more and more localized by the deep cosine potential at $\theta = \pi$. This low-energy sector very much resembles the spectrum of a particle in a box at $\omega\mu \gg 1$. We will see that there is always some (and often a lot of) IR entropy associated to such theories.

\subsection{Particle in a box}

A particle in a box does can be modeled by an infinitely repulsive $\delta$-function potential on a circle,
\bel{
  H = \alpha(L + L^{-1}) + \beta \sum_n U^n \stackrel{N \rar \infty}\longrightarrow -\frac1{2\mu}\pdder{}{\theta} + \beta \delta(\theta), \quad \beta \gg N = \delta(0).
}
There will then be $N - 1$ energy eigenstates that represent standing waves in the box, and there will be one state localized at the $\delta$-function --- the position eigenstate $\qvec{\b 1}$. The energy of the latter will be $O(\beta N)$ while the standing waves will have energies $O(1)$; therefore we may think of the standing waves as an effective low-energy description of the $\delta$-function potential on the circle.

Coarse-graining will eventually make any state become a mix that includes the high-energy localized state. From the point of view of the low-energy theory in the box, this means that after each coarse-graining, there are fewer and fewer states that do not mix with the high-energy degree of freedom. In this sense, the low-energy theory upon decimation loses degrees of freedom and has to be modeled with Hilbert spaces of ever lower dimensionality. This corresponds to the walls of the box closing in on the particle until the particle has no low-energy degrees of freedom left.

The localized state always decimates to a maximally mixed state, as can be readily checked by representing its density matrix in momentum basis and noticing that all diagonal elements are equal to $1/N$. Its entropy is thus $\log N$ in the deep IR, but since there is only one such state, its contribution to the average IR entropy is negligible.

The standing waves also contribute nontrivially to the IR entropy. Compared to the propagating waves of the free particle, a standing wave entropy is higher because the $\delta$-function forces it to have a node at $\theta = 0$, so after enough decimations it will necessarily become mixed.

The wave with $p$ humps corresponds to the state
\bel{
  \qvec{p} =  \sum_n \frac{g^{pn/2} - g^{-pn/2}}{i\sqrt{2N}}\, \qvec{g^n} \stackrel{N \rar \infty}\longrightarrow \frac1{\sqrt \pi} \int \d\theta \sin\frac{p\,\theta}2\, \qvec{e^{i\theta}}.
}
This makes sense when $0 < p < N$. The overlaps in \eqref{eq pk overlap} can be computed to be
\bel{
  \left|\qprod{\ell}p\right|^2 = \frac1{2N^2} \left|\frac{1 - g^{Np/2}}{1 - g^{- \ell + p/2}} - \frac{1 - g^{-Np/2}}{1 - g^{- \ell - p/2}}\right|^2.
}
When $p = 2q \neq 0$, we simply get
\bel{
  \left|\qprod\ell {2q} \right|^2 = \frac12 \left(\delta_{\ell,\,q} + \delta_{\ell,\, - q}\right).
}
On the other hand, when $p = 2q + 1$, we get
\bel{
  \left|\qprod\ell {2q + 1} \right|^2 = \frac 2{N^2} \left| \frac1{1 - g^{-\ell + q + 1/2}} - \frac1{1 - g^{-\ell - q - 1/2}}\right|^2.
}

For even harmonics, the entropy in the IR is simply $\log 2$. For odd harmonics, the formulae are slightly more complicated, but the sector probabilities in the deep IR are still highly peaked at $\ell = \pm(q+1)$ and $\ell = \pm(q - 1)$. Their IR entropies are thus approximately $\log 4$. The average IR entropy is then
\bel{
  \bar S\_{IR} \approx \frac32 \log 2.
}
As expected, this result is not very interesting, as it just agrees with the intuition that there is no mixing in $d = 1$. To make matters more interesting, we must go to higher dimensions. The equivalent of a particle on a circle with a $\delta$-function potential will be the famous Sinai billiard, and indeed we will see signatures of chaos there.

\subsection{The Aubry-Andr\'e model}

Before leaving the comfortable confines of $d = 1$, let us study a classically ergodic model and see how it qualitatively differs from the examples above. We consider a variant of the Aubry-Andr\'e model \cite{Aubry:1980}, i.e.~a particle in a weak quasiperiodic potential,
\bel{
  H = \alpha(L + L^{-1}) + \sum_n \left[\beta_n (U^n + U^{-n}) + i \gamma_n (U^n - U^{-n})\right] \stackrel{N\rar \infty}\longrightarrow - \frac1{2\mu} \pdder{}\theta + \beta \left[\cos(\xi \theta) + \sin(\zeta \theta)\right].
}
To ensure quasiperiodicity, $\xi$ and $\zeta$ are taken to be irrational, and we include both the sine and cosine in order to eliminate the parity symmetry and make the spectrum nondegenerate. In the discrete model, the couplings are
\bel{
  \beta_n = (-1)^n \frac{\beta}{2\pi} \frac{\xi \sin(\pi \xi)}{\xi^2 - n^2}, \quad \gamma_n = (-1)^n \frac{\beta}{2\pi} \frac{n \sin(\pi \zeta)}{\zeta^2 - n^2}.
}
This model is very similar to one in which the couplings are randomly chosen \cite{Anderson:1958vr}. The wavefunctions are localized in position space whenever the coupling is greater than the scale set by the system size ($1/N$).  The original Aubry-Andr\'e model is more sophisticated and, for almost all irrational periods, features a localization/delocalization transition in the continuum limit.\footnote{I thank Wen Wei Ho for extensive discussions on this issue.}

\begin{figure}
  \centering
  \includegraphics[width=0.4\textwidth,keepaspectratio=true]{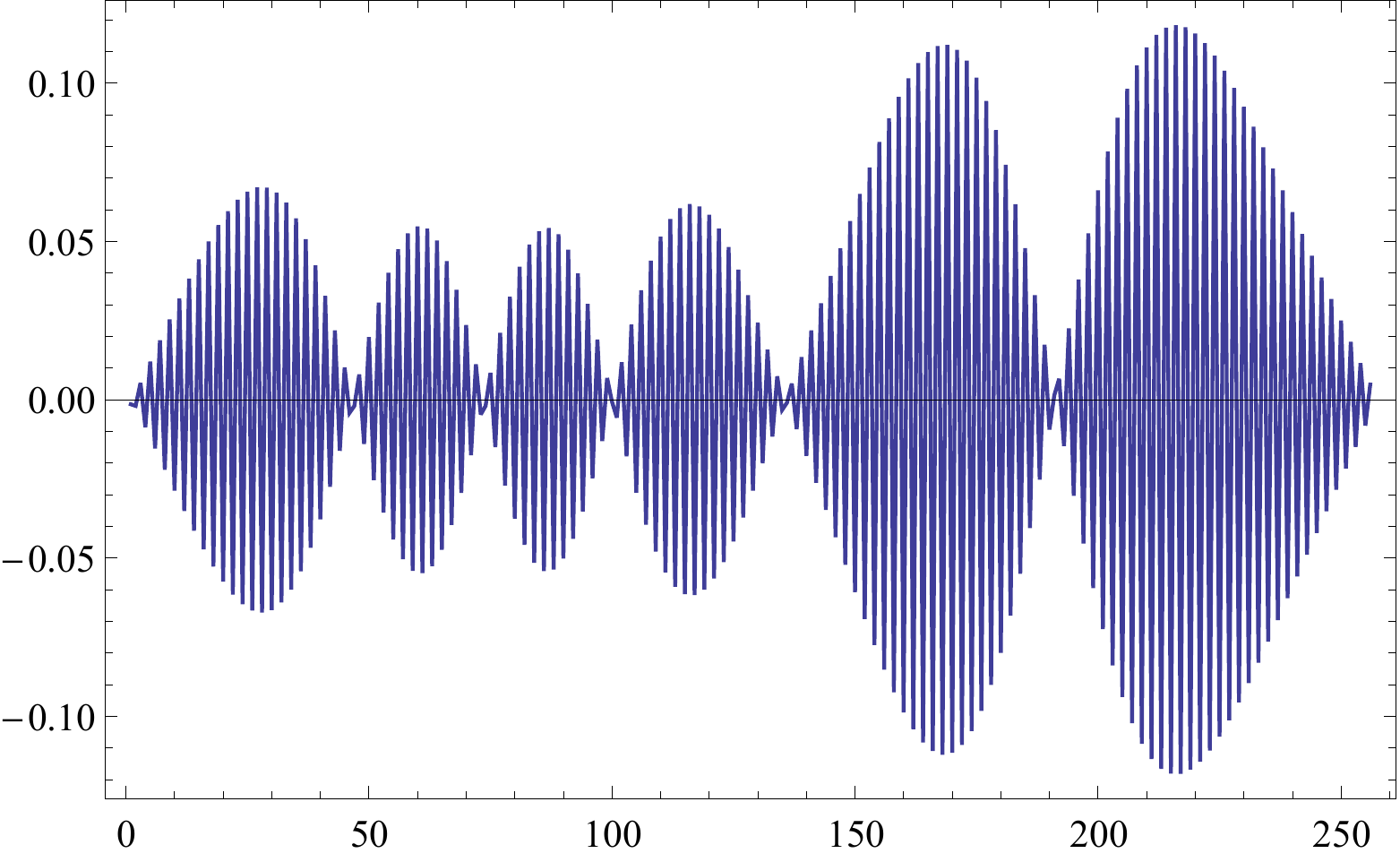} \qquad \includegraphics[width=0.4\textwidth,keepaspectratio=true]{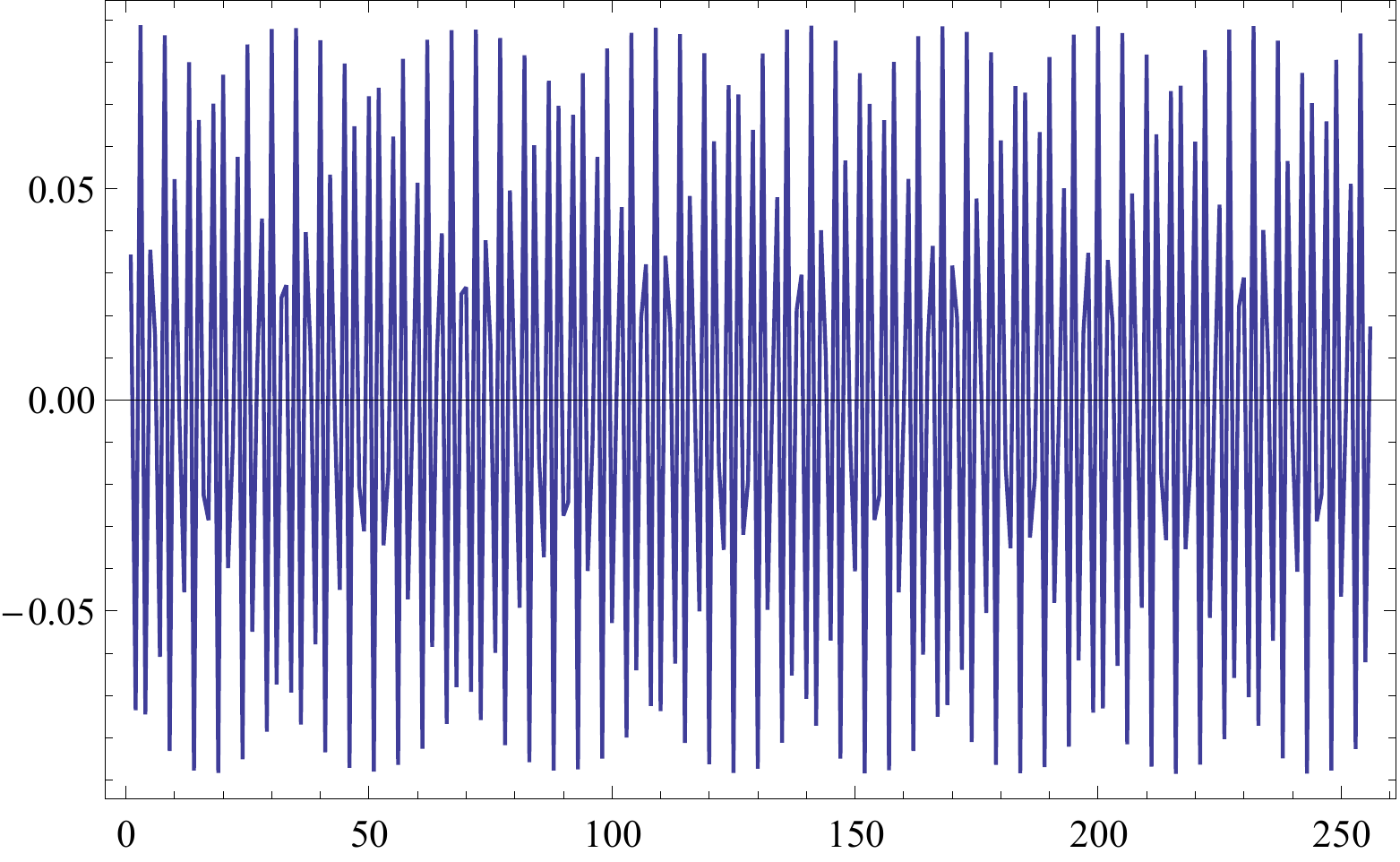}
  \includegraphics[width=0.4\textwidth,keepaspectratio=true]{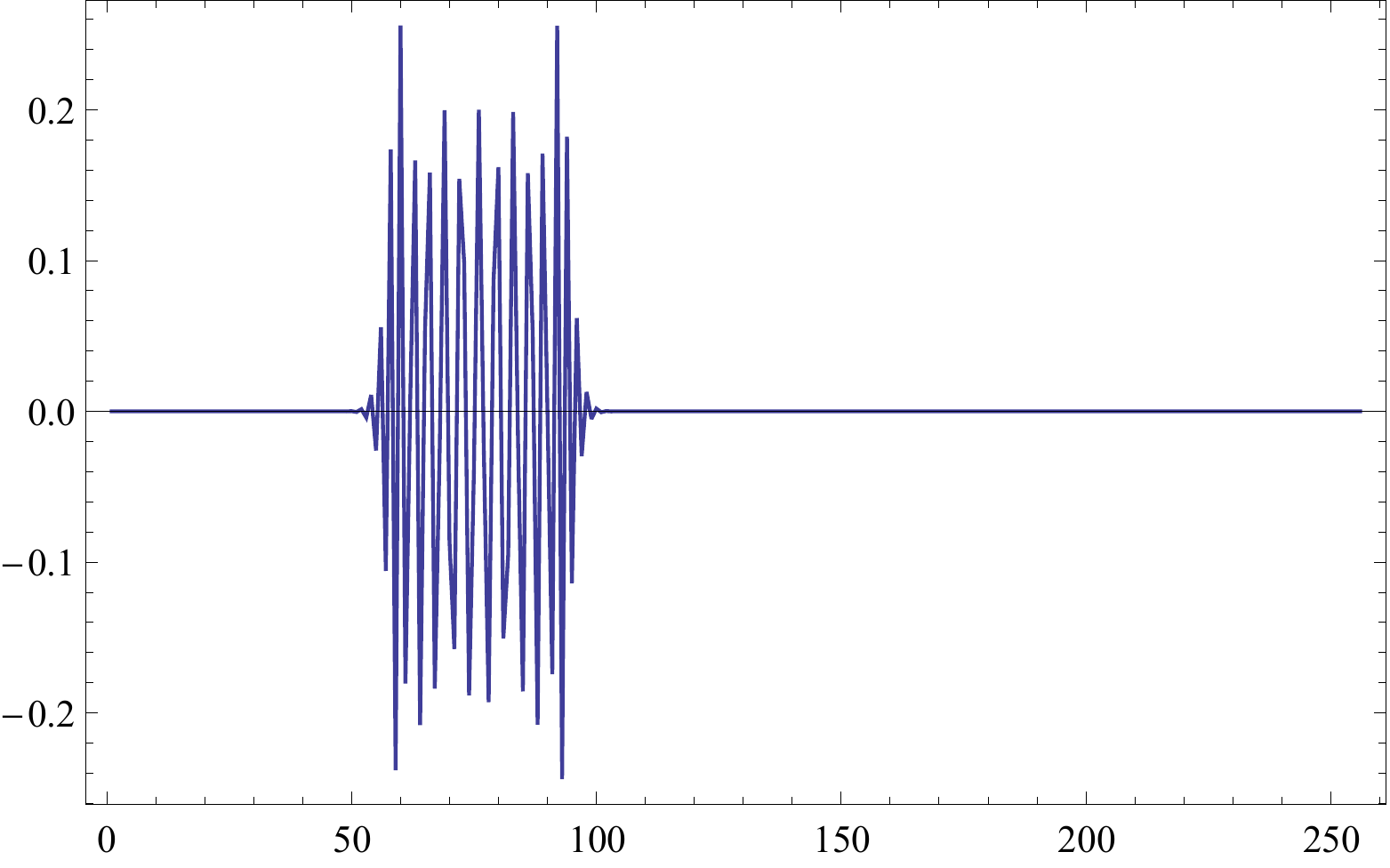} \qquad \includegraphics[width=0.4\textwidth,keepaspectratio=true]{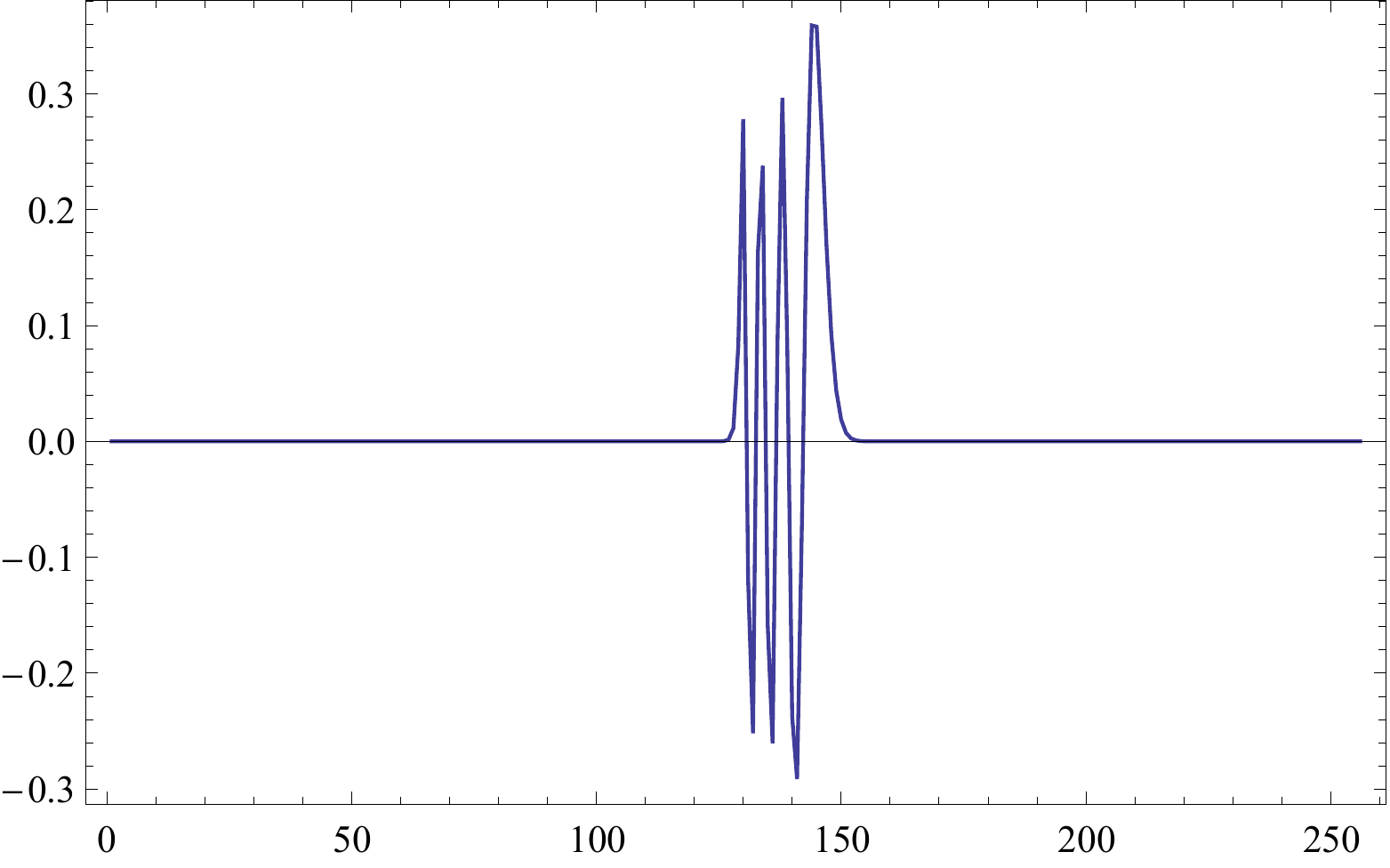}

  \caption{\small Representative energy eigenfunctions in the delocalized (above) and localized (below) phase on 256 sites in the target space. Localization of all eigenstates happens at $\beta \gtrsim 0.01 \sim 1/N$ in our model.}
  \label{fig AA wfs}
\end{figure}

We exactly diagonalize the discrete model with up to $N = 256$ sites. The onset of delocalization due to finite size effects can be seen on Fig.~\ref{fig AA wfs}, where typical eigenstates are shown at values of $\beta$ below and above $1/N$. In the localized regime the energy eigenstates are expected to have nontrivial IR entropy, just like how in the presence of the $\delta$-function the odd-numbered standing waves received extra IR entropy due to the node at $\theta = 0$. The increase in IR entropy is much more significant here as eigenstates are localized much more brutally than at just one point in position space.

\begin{figure}
  \centering
  \includegraphics[width=0.4\textwidth,keepaspectratio=true]{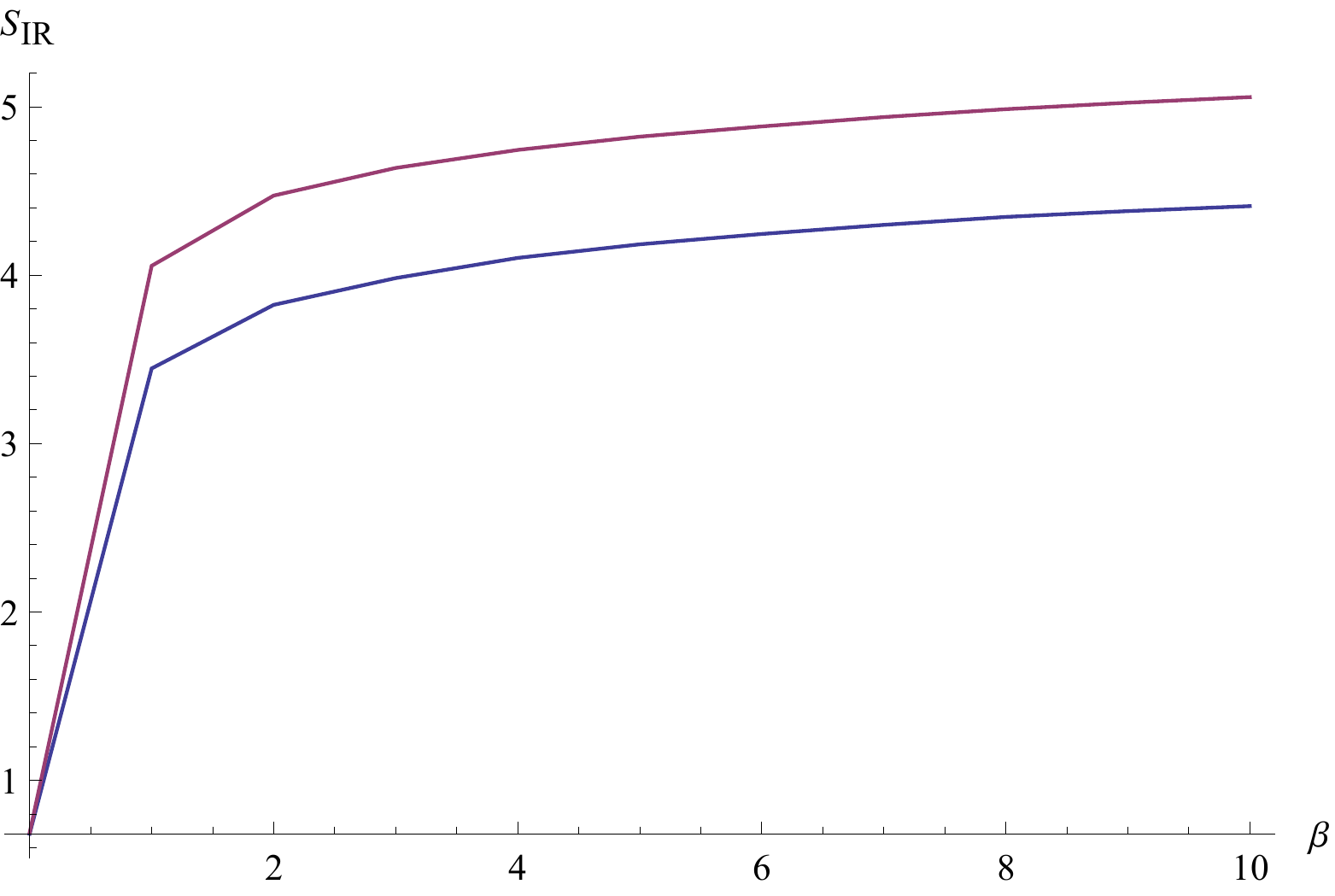} \qquad \includegraphics[width=0.4\textwidth,keepaspectratio=true]{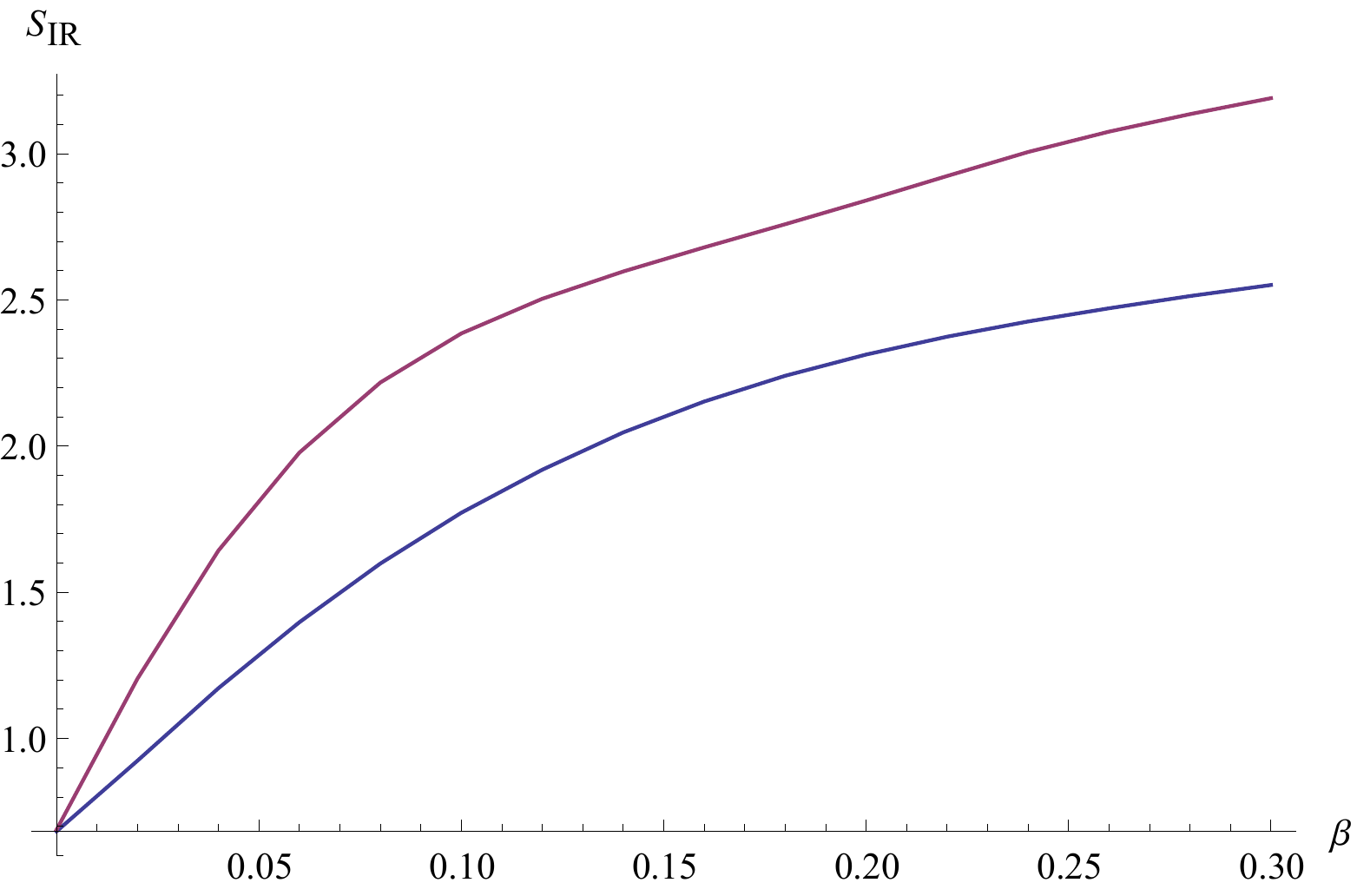}
  \caption{\small Dependence of the average IR entropy on the coupling $\beta$ for $N = 128$ (blue, lower) and $N = 256$ (purple, upper). The second graph zooms in to the area around the crossover to localization, where $\bar S\_{IR}$ is found to smoothly interpolate between the two phases. The difference between the two entropies asymptotes to $\log 2$, and each entropy asymptotes to $\log N$, its maximal possible value.}
  \label{fig AA S}
\end{figure}

The average IR entropy is straightforward to evaluate using \eqref{eq pk overlap}. The main finding is that in the delocalized regime, the IR entropy does not grow with $N$, just like in the previous examples, while in the localized regime it grows as \bel{
  \bar S\_{IR} = -s_0(\beta) +s(\beta) \log N,
}
with $s_0(\beta) > 0$. The parameters $s_0(\beta)$ and $s(\beta)$ approximately saturate the entropy bound, $s^* \approx 1$ and $s_0^* \approx 0$ at $\beta \gg 1/N$. Examples of this behavior are shown on Fig.~\ref{fig AA S}. In the next Section we will argue that this behavior of $\bar S\_{IR}$ corresponds to the lack of mixing in the underlying classical model (or, more precisely, to the lack of quantum ergodicity in the quantum model).

\subsection{A modified Sinai billiard}

Our final example is the Sinai billiard: a particle moving on a two-dimensional torus with a circular, totally reflecting obstacle. We will model this system the same way we modelled a particle in a box, by studying motion on $\Z_N \otimes \Z_N$ with several impenetrable $\delta$-functions scattered across the torus surface. A single obstacle would be enough to make the very excited states ergodic \cite{Berry:1977}, but this will provide only a $O(1/N)$ contribution to the IR entropy. The entropy will increase the more scatterers we add, especially if we break symmetries and lift degeneracies while adding them. The Hamiltonian is
\bel{
  H = -\frac1{2\mu} \left(\pdder{}{\theta_1} + \pdder{}{\theta_2}\right) + \beta \sum_{i = 1}^R \delta(\vec \theta - \vec \vartheta_i),
}
where as before $\beta \gg N$ and $\vec\vartheta_i$ are random points on the torus.

To study the influence of the scatterers, we compute the IR entropy for different numbers $R$ of randomly sprinkled $\delta$-functions on a two-torus. Each obstacle will contain a localized state which will contribute a $\log N^2$ to the sum over entropies. However, we are interested in the entropy contained in the standing waves only, as these are the eigenstates one gets when studying the Laplacian on the surface bounded by the obstacles. This is why the IR entropy we present in what follows contains an average only over these standing wave entropies. (For the particle in a $d = 1$ box, this was irrelevant, as there was only one localized state; here we may have $O(N^2)$ localized states.)

\begin{figure}
  \centering
  \includegraphics[width=0.4\textwidth,keepaspectratio=true]{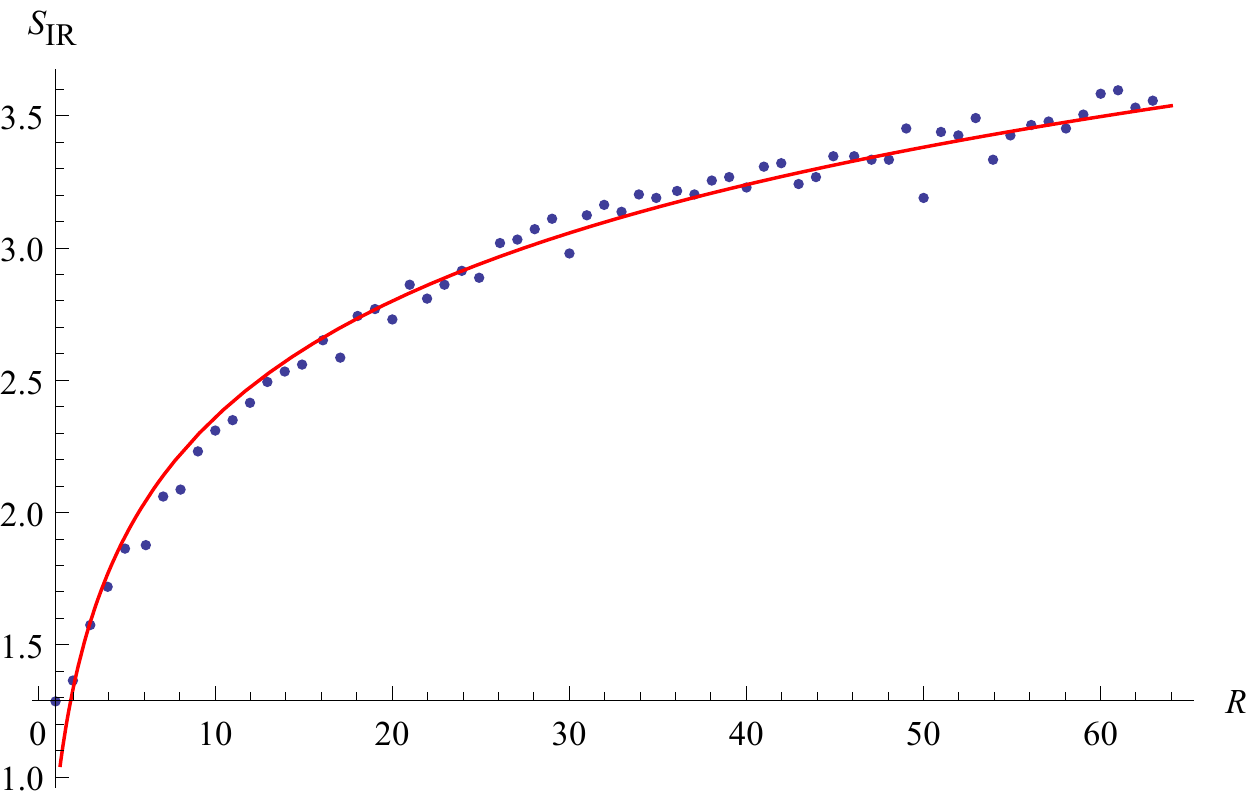} \qquad \includegraphics[width=0.4\textwidth,keepaspectratio=true]{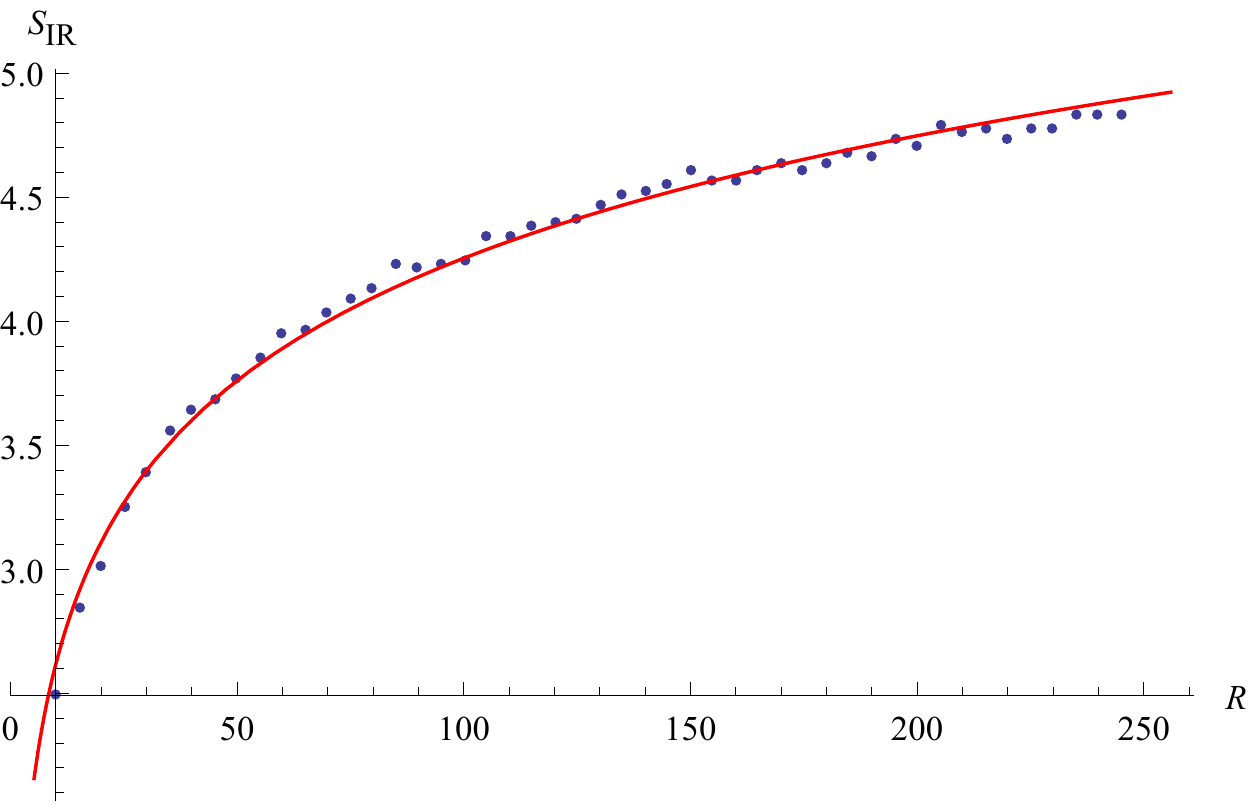}
  \caption{\small Average IR entropies of standing waves on a two-torus with $R$ scatterers, with $N^2 = 64$ (left) and $N^2 = 256$ (right). The fitted functions have form $s_0 + s_1 \log R$, and the best-fit parameters are $s_0 \sim 0.9$, $s_1 \sim 0.7$. This is just an illustrative sample; the parameters were not obtained from statistical averaging over different scatterer configurations.}
  \label{fig sinai}
\end{figure}

We find that, for $R$ scatterers, the average IR entropy of the remaining $N^2 - R$ eigenstates grows as
\bel{
  \bar S\_{IR} = s_0 + s_1 \log R.
}
These results for $N^2 = 64$ and $N^2 = 256$ are shown on Fig.~\ref{fig sinai}. The number $R$ can be construed as the size of the scattering obstacle, and we thus see that the entropy of each standing wave with a macroscopic obstacle ($R \sim N$) is of the order $\log N$ without saturating the bound $\log N^2$ until, approximately, the obstacle size fills up the entire torus at $R = N^2$. This is a striking qualitative deviation from all the previous results, where the average entropy was either $O(1)$ or essentially equal to its theoretical maximum. As we will argue in the next Section, this is the signature of quantum ergodicity.

\section{Infrared entropy and quantum ergodicity}

We now want to make connections between the calculated IR entropies and the quantum ergodicity of theories in question. Recall that a system is called quantum ergodic if, roughly, its energy eigenstates are randomly distributed, or if there is ``eigenstate thermalization.'' More precisely, this condition is fulfilled if for almost any operator, its expectation value in an energy $E$ eigenstate is equal to the average of the classical function representing the operator over the constant energy $E$ surface in classical phase space \cite{Shnirelman:1974, CdV:1985}. This happens when the underlying classical system is chaotic.

The entropy of the probability distribution \eqref{eq pk overlap} measures how much the eigenvectors of $H$ deviate from an ``ordered'' set of eigenvectors --- the eigenstates of the free particle on the underlying group. In this sense, one would expect that the larger the IR entropy, the more random the eigenstate. However, there is a duality at work here: if an eigenstate's entropy reaches its theoretical maximum $\log N$ (with $N$ the Hilbert space dimension), this state is very atypical --- it is localized in position space. Indeed, we saw in the Aubry-Andr\'e model that the localized phase has an average IR entropy that approximately saturates its maximal value. In addition, even with the cosine potential, we saw that the entropy was a monotonically increasing function of $\mu\omega$. If we took this coupling to infinity, all eigenstates would become localized in the bottom of the cosine potential, and indeed we would retrieve $\bar S\_{IR} = \log N$.

This situation can be compared to that in quantum field theories on a lattice when both sides of a strong-weak coupling duality are tractable. Examples include the Ising chain in a transverse field and the Kitaev toric code in two spatial dimensions. The strong coupling regimes in these two theories would be called ``ordered'' and ``confined,'' respectively; at weak coupling the regimes are disordered and deconfined/topological, which are just order and confinement in the dual basis of states. At strong coupling, the ground state entanglement entropy of a region is zero; at weak coupling, the entropy is the maximal possible one consistent with locality and global constraints. We see that in QM the same story plays out, with ``ordered'' replaced by ``free'' (or ``localized in momentum space'') and ``disordered'' replaced by ``localized in position space,'' and with the IR entropy playing the role of the ground state entanglement entropy.

We are thus led to expect that a theory is quantum ergodic when its average IR entropy diverges in the continuum/thermodynamic limit ($N \rar \infty$) --- but only when it does not diverge maximally fast. In particular, we may speculate that the most chaotic classical theories give $\bar S\_{IR} = s \log N$ upon quantization, with $s$ being a potentially universal number near $1/2$. Deviations from this conjectural baseline would then reflect the existence of nonergodic states like quantum scars \cite{Heller:1984}.

Ideas like these are likely not new to the cognoscenti of quantum chaos. The novelty of our approach is the exclusive usage of groups as target spaces. One benefit of this new approach is that free particle eigenstates (both the position and the momentum ones) can be defined with reference just to the group structure, and thus provide a privileged set of eigenstates with respect to which the overlaps in \eqref{eq pk overlap} are to be measured.

A different but complementary benefit is that the notion of algebraic decimation naturally gives rise to the entropy of sector probabilities as a measure of long-time behavior (and, therefore, of  quantum ergodicity). Moreover, performing a few decimations already allows us to compute an entropy function via eq.~\eqref{eq pk vev}, giving us a natural coarsening of the microscopic overlap function that may still be a useful diagnostic.

\section{Discussion}

Not only is our renormalization group of quantum mechanics not a group, it also does not renormalize any infinity. Nevertheless, we have argued that it is still a very useful operation to consider: it formalizes the Wilsonian idea of flowing to longer times and allows us to understand the fixed points of quantum mechanics in the deep infrared. We have shown that for any quantum mechanics defined on a flat group manifold, the fixed point is not another quantum mechanics --- instead, it is a classical theory of probability. The entropy of this infrared probability distribution is a natural diagnostic of quantum ergodicity while naturally sharing many parallels with other entropies from QFT.

The following points are worth emphasizing at the close of this paper:
\begin{enumerate}
  \item The group structure was crucial to defining and interpreting the algebraic decimation. Many quantum mechanical problems lack an apparent underlying group structure, including some integrable ones like the Calogero model. Perhaps a group structure can be uncovered in each such problem, just how we have interpreted a particle in a box as a particle on a torus with a totally reflecting barrier.
  \item We have not addressed the infrared fate of particles moving on nonabelian groups. For positively curved groups, this is a straightforward, if tedious, extension of the methods presented here. For hyperbolic spaces (which are of great interest on their own), a few more tools seem to be needed, and this analysis will be presented in a separate publication \cite{GARS}.
  \item One of our main results is that there exists no effective quantum mechanics that governs the evolution in the infrared. Our analysis was valid for all Hilbert space dimensions. However, perhaps when this dimensionality is large, an approximate, \emph{statistical} description of the IR evolution can be defined. Understanding this possibility would be a valuable goal.
  \item Our results were all derived for Hilbert spaces whose dimensions were powers of two. Extending these to other powers requires a straightforward modification of the decimation procedure so as to involve different powers of original position generators. Amusingly, if $N$ is prime, there is only one decimation needed before reaching the classical probability regime --- independently on the size of $N$!
  \item A single Majorana fermion is an operator $\chi$ that satisfies $\chi^2 = \1$. (Its fermionic nature is only apparent when other fermions are around.) In the context of our work, we can recognize the Majorana algebra as the Abelian algebra that quantum mechanics on $\Z_2$ reduces to after one decimation. In other words, a single Majorana fermion shoud be viewed as an Ising spin for which we can only measure spin along one direction. Its natural Hilbert space is two-dimensional, and the possible set of states of one Majorana are all classical probability distributions on two elements. Generalizations of Majorana fermions are obtained by decimating any $\Z_N$ group when $N$ is prime.
  \item Degeneracy of the spectrum can lead to ambiguity in the infrared entropy. We manually fixed this issue when it arose, but a more systematic treatment might be possible.
  \item We end by simply emphasizing the most important question that this paper fails to address: the origin of random matrix universality in quantum chaotic systems. We have argued that going to long times does \emph{not} give rise to an effective Hamiltonian that is a random matrix. A different kind of flow and universality potentially await to be uncovered.
\end{enumerate}

\section*{Acknowledgments}

It is a great pleasure to thank Jaume Gomis, Guy Gur-Ari, Wen Wei Ho, Raghu Mahajan, Xiao-Liang Qi, Grant Salton, Steve Shenker, and Sho Yaida for useful conversations. The author was supported by a Stanford Graduate Fellowship and has benefited from the hospitality of Perimeter Institute while a part of this work was done. Research at Perimeter Institute is supported by the Government of Canada through Industry Canada and by the Province of Ontario through the Ministry of Economic Development \& Innovation.

\end{document}